\documentclass[iop]{emulateapj}
\usepackage{apjfonts}
\usepackage[usenames]{color}
\usepackage{hyperref}

\newcommand{\cM}{{M}}

\newcommand{\mybibitem}[3]{\bibitem[{#1}({#2})]{#3}}
\newcommand{\mybibthree}[4]{\bibitem[{#2}({#3}){#1}]{#4}}
\newcommand{\picplace}[1]{\vbox{\hrule\@height 0.4pt\@width\hsize
\hbox to\hsize{\vrule\@width 0.4pt\@height#1\hfil
\vrule\@width 0.4pt\@height#1}\hrule\@height 0.4pt\@width\hsize}}

\slugcomment{To appear in ApJ, {\bf\textit{833}}, 8 (2016)}

\shorttitle{Evolution of the Globular Cluster Luminosity Function}
\shortauthors{Paul Goudfrooij \& S. Michael Fall}

\begin{document}


\title{Evolution of the Mass and Luminosity Functions of Globular Star Clusters}   


\definecolor{MyBlue}{rgb}{0.3,0.3,1.0}

% The following is for the preprint version
%\author{Paul Goudfrooij and S.\ Michael Fall}
%\affil{Space Telescope Science Institute, 3700 San Martin
%  Drive, Baltimore, MD 21218, USA}
%\email{{\textcolor{MyBlue}{\href{mailto:goudfroo@stsci.edu}{goudfroo@stsci.edu}}},
%      \textcolor{MyBlue}{fall@stsci.edu}} 

%\and

%\author{Michele Trenti}
%\affil{School of Physics, The University of Melbourne, VIC 3010, Australia}
%\email{\textcolor{MyBlue}{mtrenti@unimelb.edu.au}}

% The following is for the emulateapj version
\author{Paul Goudfrooij and S.\ Michael Fall \vspace*{0.3mm}}
\affil{Space Telescope Science Institute, 3700 San Martin
  Drive, Baltimore, MD 21218, USA; {\color{MyBlue}goudfroo@stsci.edu}}
%\email{goudfroo@stsci.edu}

%\altaffiltext{1}{Based on archival observations with the NASA/ESA {\it Hubble
%    Space Telescope}, obtained at the Space Telescope Science
%  Institute, which is operated by the Association of Universities for
%  Research in Astronomy, Inc., under NASA contract NAS5-26555} 

%% Notice that these authors may have alternate affiliations, which


\begin{abstract}
We reexamine the dynamical evolution of the mass and luminosity functions of
globular star clusters (GCMF and GCLF).  Fall \& Zhang
(2001, hereafter FZ01) showed that a power-law MF, as commonly seen
among young cluster systems, would evolve by dynamical processes over a
Hubble time into a peaked MF with a shape very similar   
to the observed GCMF in the Milky Way and other galaxies. 
To simplify the calculations, the semi-analytical FZ01 model adopted the 
``classical'' theory of stellar escape from clusters, and
neglected variations in the $\cM/L$ ratios of clusters. 
Kruijssen \& Portegies Zwart (2009, hereafter KPZ09)  
modified the FZ01 model to include ``retarded'' and mass-dependent stellar 
escape, the latter causing significant $\cM/L$ variations. KPZ09 asserted that their
model was compatible with observations whereas the FZ01 model was not. 
We show here that this claim is not correct; the FZ01 and KPZ09 models fit
the observed Galactic GCLF equally well.  
We also show that there is no detectable correlation between $\cM/L$ and
$L$ for GCs in the Milky Way and Andromeda galaxies, in contradiction with the
KPZ09 model.    
Our comparisons of the FZ01 and KPZ09 models with observations can be
explained most simply if stars escape at rates approaching the classical limit
for high-mass clusters, as expected on theoretical grounds.    
\end{abstract}





%----------------------------- intro section ------------------------------

\section{Introduction}              \label{s:intro}
One of the most remarkable properties of globular cluster (GC)
systems is the similarity of their luminosity functions from one
galaxy to another. These have bell-like shapes and are often modeled
as log-normal distributions of luminosities or, equivalently, Gaussian
distributions of magnitudes (see, e.g., \citealt{harr91}).  
In contrast, the luminosity functions of young cluster systems
are always found to be power laws, $\phi(L) = dN/dL \propto
L^{\alpha}$ with $\alpha \approx -2$ (\citealt{vdblaf84,elsfal85,chrsch88};
see \citealt{whit+14} for a recent comprehensive study of 20 star-forming
galaxies).  

The mass functions of cluster systems have greater dynamical
significance than their luminosity functions.  For systems of old, coeval 
clusters, variations in the mass-to-light ($\cM/L$) ratios are relatively
small, and the luminosity function is a good proxy for the mass
function.  Thus, the mass function of globular clusters (GCMF) must
have nearly the same bell-like shape as the luminosity function
(GCLF).  However, systems of young clusters in the process of formation
have wide spreads in $\cM/L$, and the mass function must be determined
separately from the luminosity function, usually by estimating the masses,
ages, and reddenings of individual clusters in the sample from multi-band
photometry. (See \citealt{fall06} for the mathematical relations between the
mass, luminosity, and age distributions of cluster systems.)  Studies of
this kind show that the mass functions of young cluster systems 
are power laws, $\psi(M) = dN/dM \propto M^{\beta}$ with $\beta \approx
-2$ (\citealt{zhafal99,falcha12}, and references therein). As a consequence,
larger systems of young clusters will contain more massive clusters, reaching $M
\sim 10^6 \;M_{\odot}$ or even $\sim 10^7 \; M_{\odot}$ in some cases
\citep{chan+10}. The most massive of these clusters are often referred
to as ``young GCs''. 

There are two possible explanations for the radically different mass
functions of young and old cluster systems: either (1) the process of
cluster formation, and hence the initial cluster mass function, differed in the
distant past from the present, or (2) the mass function of clusters evolves
by dynamical processes from an initial power law into a bell-shaped
distribution.  The second possibility---the one we examine in this
paper---has been explored at various levels of approximation over the years
\citep{falree77,gneost97,baum98,vesp98,prigne08}. 
We focus here on the semi-analytical model for the evolution of the mass
function developed by \citet[][hereafter FZ01]{falzha01}.   

In the FZ01 model, clusters are tidally limited at the pericenters of their
galactic orbits and are disrupted by the gradual escape of stars driven by a
combination of internal two-body relaxation and external gravitational
shocks.  For most clusters, shocks are relatively weak, and relaxation is the
dominant disruption mechanism. According to the ``classical'' theory of
relaxation-driven stellar escape, as formulated by \citet{spit87} and others,
the mass $M$ of a tidally limited cluster decreases at a nearly constant rate:
$dM/dt  = \mu$ with $\mu \propto \rho_{\rm h}^{1/2}$, where $\rho_{\rm h} = 3 M
/ (8 \pi r_{\rm h}^3)$ is the mean density within the half-mass radius $r_{\rm
  h}$ of the cluster.  
FZ01 showed that, in this approximation, the evolving GCMF at any time
$\psi(\cM,\,t)$ is related to the initial GCMF $\psi_0(\cM)$  
by  $\psi(\cM,\,t) = \psi_0(\cM\,+\,\mu\,t)$.  This has a characteristic bend
or peak at $M \sim \mu\, t$ and the limiting forms $\psi(M, t) =
\psi_0(\mu\,t)$, independent of $M$ for $M \ll  \mu\,t$, and $\psi(M,
t) = \psi_0(M)$, independent of $t$ for $M \gg \mu\,t$.  

FZ01 compared their model to the observed GCLF in the Milky Way in the
approximation of constant $M/L$ and found excellent agreement.  In particular,
they showed both theoretically and observationally, for the first time, that
the GCMF and GCLF are approximately constant for $M \la 10^5 \; M_{\odot}$ and 
$L \la 10^5 \; L_{\odot}$, in stark contradiction to the then-standard practice of
fitting log-normal distributions to the data.  This behavior of the GCMF at
small $M$ and GCLF at small $L$ is strong evidence for the late disruption of
clusters by internal two-body relaxation. 
Furthermore, \citet{goud+04,goud+07} showed that the predicted time
dependence of the FZ01 model is consistent with the observed luminosity
functions of intermediate-age (3\,--\,4 Gyr old) cluster systems, and
\citet{chan+07}, \citet{mclfal08}, and \citet{goud12} showed that the predicted
density dependence agrees well with the observed mass function for subsamples of
clusters defined by different ranges of density. 

The approximations in the FZ01 model were made for simplicity and to
highlight the main physical processes that shape the GCMF and GCLF.  In
particular, the adoption of classical evaporation and the neglect of $M/L$ 
variations are not essential features of the model. Both approximations were
in standard use at the time (2001) to describe the evolution of individual
clusters. The novel feature of the FZ01 model was to show how the evolution
of the masses of individual clusters could be combined analytically into the
evolution of the mass function of a cluster system. 

At a higher level of approximation, the stellar escape rate is modified by
the fact that some of the stars that are scattered into unbound orbits may be
scattered back into bound orbits before they have reached the tidal boundary
and escaped from a cluster.  In this case, often called ``retarded''
evaporation, the escape rate has a weak dependence on the crossing time $t_{\rm
  cr}$ in addition to the stronger dependence on the relaxation
time $t_{\rm rlx}$.  For tidally limited, low-mass clusters (initial
  masses $M_0 \la 10^{5} \;  M_{\odot}$), the evolution can be approximated
by $dM/dt \propto M/t_{\rm dis}$ with $t_{\rm dis} \propto M^{\gamma}$ and
$\gamma \approx 0.7$, rather than $\gamma$ = 1 for classical evaporation
\citep{fukheg00,baum01,baumak03,lame+10}. For high-mass clusters, however, the
retarded evaporation rate must approach the classical rate (in the
limit $t_{\rm cr}/t_{\rm rlx} \rightarrow 0$; see Section \ref{sub:gamma}).     

Two-body relaxation will also cause low-mass stars within a cluster to gain
energy and escape faster than high-mass stars, thus reducing the average
$\cM/L$ of the remaining stars over and above the fading caused by
stellar evolution alone.  As a result of this effect, in a coeval population
of clusters (such as GCs), there should be a positive correlation between $M/L$ 
and the mass or luminosity of clusters, because those
that have smaller relaxation times will have lost larger fractions of their
initial mass.  Such variation in $M/L$ was neglected in the FZ01 model for two
reasons:\ it is difficult to predict reliably from theory, and it appeared 
from observations at the time to be weak or non-existent
\citep{mcla00}. 

\citet[][hereafter KPZ09]{krupor09} modified the FZ01 model to include
retarded evaporation and a variable $\cM/L$ ratio.  In particular, they
assumed that the rate of mass loss from clusters is given by $dM/dt \propto
M/t_{\rm dis}$ with $t_{\rm dis} \propto M^{\gamma}$ and $\gamma = 0.7$
for clusters of all masses. Furthermore, to calculate the escape
rates of stars of different masses and hence the variation in $M/L$ of
clusters, they employed a semi-analytical model developed by
\citet[][hereafter K09]{krui09} that includes several questionable
  assumptions and parameter choices. 
KPZ09 argued that their model is a significant improvement on the FZ01 model,
both in terms of its theoretical validity and in terms of its ability to fit the
observed GCLF in the Milky Way and other galaxies.  

Our main purpose in this paper is to demonstrate that the KPZ09 criticisms of
the FZ01 model are not correct. Thus motivated, we also show that the
variation in $M/L$ with $M$ or $L$ predicted by the KPZ09 model is much
stronger than that allowed by observations. Furthermore, we show that the
parameter values required for the KPZ09 model to fit the observed GCLF and the
observed $\cM/L$ vs.\ $L$ relation are mutually exclusive. 
We emphasize that we do not dispute the general physical principles
underlying retarded evaporation and $M/L$ variations.  The results of this
paper indicate, however, that the specific implementation of these effects in
the KPZ09 model exaggerates their importance. 
We find that retarded evaporation and $M/L$ variations can be
neglected for clusters massive enough to survive for a Hubble time of
dynamical evolution. 
Therefore, for most practical purposes, the benefits of including
these effects are largely offset by the increased complexity of the
KPZ09 model relative to the FZ01 model. 

This paper is organized as follows. 
In Section~\ref{s:GCLFfits}, we compare the FZ01 and KPZ09 models
with the observed GCLF in the Milky Way, and we determine the 
best-fitting values of their parameters including the characteristic
dissolution timescale.  
Section~\ref{s:dyn} presents our search for $\cM/L$ variations of the kind 
predicted by the KPZ09 model in a large compilation of recent dynamical 
measurements of GC masses. 
In Section~\ref{s:Tdiss}, we compare the dissolution timescale required by 
the KPZ09 model to match the observed GCLF with the one required by the 
absence of observed $\cM/L$ variations. 
Section~\ref{s:disc} interprets the results from Sections~\ref{s:dyn} and
\ref{s:Tdiss} along with the observed stellar mass functions in Galactic GCs in
terms of key properties and assumptions of the K09 model. Finally, we summarize
our conclusions in Section~\ref{s:conc}.

\section{Comparisons of Models with the Observed GCLF}  
\label{s:GCLFfits}

In this Section, we compare the FZ01 and KPZ09 models with the observed GCLF
in the Milky Way. We first derive analytical expressions for the evolving
GCMFs assuming that the clusters are tidally limited and that stellar escape
driven by two-body relaxation is the primary disruption mechanism. 
We then convert these GCMFs at an age of 12 Gyr into GCLFs adopting $\cM/L$ =
constant for the FZ01 model and the $\cM/L$ vs.\ $L$ relation derived by KPZ09
for their model.  

\subsection{Derivation of Model GCMFs}
\label{sub:GCMFs}

In both models considered here, the mass-loss rate of an
individual cluster takes the form 
\begin{equation}
dM/dt \; \equiv \; -M/t_{\rm dis} \; = \; -(\mu/\gamma) \; M^{1-\gamma}\mbox{,}
\label{eq:dMdt}
\end{equation}
where $t_{\rm dis}$ is the dissolution timescale and $\mu$ and 
$\gamma$ are constants. This integrates to 
\begin{equation}
\cM\,(t) \; = \; (\cM_0^\gamma - \mu \, t)^{1/\gamma}\mbox{,}
\label{eq:massevol}
\end{equation}
where $\cM_0$ is the initial cluster mass. 
These formulae are intended to represent smooth averages over the abrupt
changes in mass caused by the escape of individual stars and over at least one
full orbit of the cluster around its host galaxy. For $\gamma = 1$,
equation~(\ref{eq:dMdt}) describes the classical mass-loss rate $\mu$ for
stellar escape driven by internal two-body relaxation from a
tidally limited cluster (\citealt{spit87} and references therein). This is
the formula adopted in the FZ01 model. For $\gamma < 1$, 
equation~(\ref{eq:dMdt}) approximates the corresponding mass-loss rate for
retarded evaporation in relatively low-mass clusters
for which crossing times are significant fractions of relaxation times
\citep{fukheg00,baumak03}. The KPZ09 model
assumes $\gamma = 0.7$ for clusters of all masses.  

FZ01 showed that the evolution of the mass function $\psi\, (\cM,\,t)$ of a
cluster system could be derived from the evolution of the masses $\cM(t)$ of
individual clusters through a continuity equation. This approach yields 
\begin{equation}
\psi\, (\cM,\,t) \; = \; \left({\partial \cM_0}/{\partial \cM}\right)_t \; 
 \psi_0 \, (\cM_0) \; = \; \left({\cM}/{\cM_0}\right)^{\gamma-1} \;
 \psi_0\,(\cM_0)\mbox{,} 
\label{eq:MF}
\end{equation}
where $\psi\,(\cM_0) = \psi\,(\cM_0, 0)$ is the initial mass function
(at $t = 0$). In this expression, the initial mass $\cM_0$ must be regarded
as a function of the current mass $\cM$ and current time $t$ as given by
inverting equation~(\ref{eq:massevol}). Following FZ01, we adopt a
\citet{sche76} initial mass function 
\begin{equation}
\psi_0 \, (\cM_0) = A \; \cM_0^{\beta} \; \exp\,(- \cM_0/\cM_{\rm c})\mbox{,}
\label{eq:schech}
\end{equation}
with adjustable parameters $A$, $\beta$, and $M_{\rm c}$. This
function has a power-law shape with exponent $\beta$ below the bend at 
$\cM_{\rm c}$ to mimic the observed mass functions of 
young cluster systems, and it has an exponential decline above $\cM_{\rm
  c}$ as suggested by the observed tail of the GCMF at
$\cM \ga 10^6 \; M_{\odot}$.
Inserting equation~(\ref{eq:schech}) into equation~(\ref{eq:MF}) then yields
the evolving GCMF:
\begin{eqnarray}
\psi\,(\cM,\,t) & = & A\;\cM^{\gamma - 1} \; (\cM^{\gamma} + \mu \,
  t)^{(\beta - \gamma + 1)/\gamma} \nonumber \\
 & & \qquad \qquad \qquad \mbox{} \times \exp\left[- \, (\cM^{\gamma}
   + \mu \, t)^{1/\gamma}/\cM_{\rm c} \right]\!\!\mbox{.} 
\label{eq:finMF}
\end{eqnarray}
This function has a bend at $\cM \sim (\mu\,t)^{1/\gamma}$; for lower $\cM$,
it behaves as $\psi\,(\cM,\,t) \propto \cM^{\gamma-1}$, characteristic of
dissolution by two-body relaxation, while for higher $\cM$, it behaves as
$\psi\,(\cM,\,t) \propto \cM^{\beta}\,\exp\,(-\cM/\cM_{\rm c})$, independent
of $\gamma$. Thus, we expect only minor differences in the shapes of the GCMF
between $\gamma = 1$ (FZ01 model) and $\gamma = 0.7$ (KPZ09 model).  

The GCMF derived above is strictly valid only in the idealized case
that all clusters in the GC system dissolve at the same rate $\mu$. In
reality, clusters with different internal densities, determined mainly by the
galactic tidal field, will dissolve at different rates $\mu_i$. In that
case, equation~(\ref{eq:finMF}) can be re-interpreted as the
probability density that an individual cluster with evaporation rate $\mu$ 
has a mass $M$ at an age $t$. The GCMF of a system of $\cal{N}$ coeval
clusters is then the sum of the individual probability densities: 
\begin{eqnarray}
\psi\,(M,\,t) & = & \sum_{i = 1}^{\cal{N}} \; A_i\;\cM^{\gamma - 1} \; 
  (\cM^{\gamma} + \mu_i \, t)^{(\beta - \gamma + 1)/\gamma} \nonumber \\ 
 & & \qquad \qquad \qquad \mbox{} \times \exp\left[- \,
    (\cM^{\gamma} + \mu_i \, t)^{1/\gamma}/\cM_{\rm c} \right]\!\!\mbox{.} 
\label{eq:multi-mu-MF}
\end{eqnarray}
Here $\beta$ and $M_{\rm c}$ are assumed to be the same for all clusters in
the GC system, and the normalization factors $A_i$ must be chosen such that
the integral over all $M$ is unity for each term in the sum. 

% Discuss derivation of mu_i for classical and retarded evaporation
We now relate the evaporation rate $\mu$ of a cluster to its mean
density $\rho_{\rm h}$ within the half-mass radius $r_{\rm h}$ as
follows. The $N$-body simulations of \citet{baum01} and \citet[][hereafter
BM03]{baumak03} showed that the dissolution time $t_{\rm dis}$ of a tidally
limited cluster can be approximated by 
\begin{equation}
t_{\rm dis} \propto t_{\rm rlx} \; (t_{\rm cr}/t_{\rm rlx})^{1-\gamma}
\propto M^{\gamma} \rho_{\rm h}^{-1/2}\mbox{,}
\label{eq:t_dis_BM03}
\end{equation}
where $t_{\rm rlx} \propto M^{1/2}\, r_{\rm h}^{3/2}$ is the half-mass
relaxation time and $t_{\rm cr} \propto M^{-1/2}\, r_{\rm h}^{3/2}$ is
the half-mass crossing time. The parameter $\gamma < 1$ in
equation~(\ref{eq:t_dis_BM03}) is the same as that in
equations~(\ref{eq:dMdt})\,--\,(\ref{eq:multi-mu-MF}) and measures the
deviation of the dissolution time from the formula $t_{\rm dis}
\propto t_{\rm rlx}$ for classical evaporation.\footnote{BM03 used
  the notation $x$ instead of $\gamma$ in equation~(\ref{eq:t_dis_BM03}). 
  The difference between the two is negligible: $x$ and $\gamma$ were
  derived by evaluating $t_{\rm dis}$ as functions of $N$ (number of particles)
  and the cluster mass $M$, respectively \citep[see][]{lame+10}.}  
The extra factor of $(t_{\rm cr}/t_{\rm rlx})^{1-\gamma}$ for retarded
evaporation comes about because unbound stars take a finite time,
proportional to $t_{\rm cr}$, to cross a cluster before escaping from it.
During that time, some of the unbound stars will be scattered back
into bound orbits within the cluster, thus retarding its evaporation. 
From equations (\ref{eq:t_dis_BM03}) and (\ref{eq:dMdt}), we obtain    
\begin{equation}
\mu \propto M^{\gamma}/t_{\rm dis} \propto \rho_{\rm h}^{1/2}\mbox{,}
\label{eq:mu}
\end{equation}
independent of $\gamma$ for both classical and retarded
evaporation.\footnote{To avoid confusion, we note that $\mu_{\rm ev}$ in
  \citet{mclfal08} is related to our $\mu$ by $\mu_{\rm ev} =
  (\mu/\gamma)\;M^{1-\gamma}$.}

The evaporation rate of a cluster can also be expressed in terms of its mean
density $\rho_{\rm t}$ within the tidal radius $r_{\rm t}$. This density is 
determined largely by the tidal field at the pericenter $R_{\rm p}$ of the
orbit of the cluster within its host galaxy: $\rho_{\rm t} \propto G^{-1}\,
\left(V_{\rm c,\,p} / R_{\rm p}\right)^2$, where $G$ is the
gravitational constant, and $V_{\rm c,\,p}$ is the galactic circular
velocity at $R_{\rm p}$ \citep{king62,inna+83}. \citet{mclfal08} showed that,
while the densities $\rho_{\rm h}$ and $\rho_{\rm t}$ of GCs in the Milky Way
span four or five orders of magnitude, the quantity $(\rho_{\rm t} / \rho_{\rm
  h})^{1/2}$ varies by less than a factor of two. Thus, to a good approximation,
we can rewrite equation~(\ref{eq:mu}) in the form  
\begin{equation}
\mu \propto \rho_{\rm t}^{1/2} \propto V_{\rm c,\,p} / R_{\rm p}\mbox{.}
\label{eq:mu_V/R}
\end{equation}
In an idealized static and spherical galactic potential, the pericenters of all
orbits remain fixed, and $\rho_{\rm t}$ and hence $\mu$ are constants of motion. 

Both FZ01 and KPZ09 computed the evaporation rates of clusters on different
orbits from equation~(\ref{eq:mu_V/R}) and then summed over a realistic
distribution of orbits to determine the mass function $\psi(M,\,t)$ of a GC
system from equation~(\ref{eq:multi-mu-MF}) or its integral equivalent.
However, as \citet{mclfal08} pointed out, the only role of the orbits in this
calculation is to determine the cluster densities, $\rho_{\rm h}$ or $\rho_{\rm
  t}$, a step that can be eliminated by computing $\mu$ directly from the
observed values of $\rho_{\rm h}$ or $\rho_{\rm t}$.  Because tidal radii are 
notoriously uncertain, evaporation rates are much more robust when computed
from $\rho_{\rm h}$ than from $\rho_{\rm t}$.  This is the approach we take in
this paper. Another simplification noted by \citet{mclfal08} is that
the mass function $\psi(M,\,t)$ of a GC system computed from equation
(\ref{eq:finMF}) with the median value of $\mu$ is very similar to that
computed from equation (\ref{eq:multi-mu-MF}) with a realistic distribution of
$\mu$ (see also KPZ09).  We refer to the former as single-$\mu$ models and the
latter as multiple-$\mu$ models.  In this paper, we present results for both
types of models, confirming their similarity.

The model GCMFs described above assume that evaporation by
two-body relaxation is the dominant disruption mechanism. As such, they neglect 
the effects of stellar evolution and gravitational shocks. Mass loss by stellar
evolution is dominated by supernovae and strong winds of massive stars in the
first few $10^8$ years. This material is assumed to escape from clusters of all
masses, thus leaving the shape of the GCMF unchanged. Meanwhile, for
surviving GCs in the Milky Way, FZ01 showed that mass loss due to gravitational
shocks is generally much weaker than that due to two-body relaxation for clusters
with masses below the peak of the GCMF (see also \citealt{gneost97,dine+99}). 
Moreover, the rate of mass loss by gravitational shocks depends only on the
densities $\rho_{\rm h}$ of clusters, not their masses, which preserves the
shape of the GCMF in the sense that both $\psi$ and $M$ are simply rescaled by
time-dependent factors (see FZ01).

\subsection{Fits of Model GCLFs to Observations}
\label{sub:GCLFfit}

Fitting equation~(\ref{eq:finMF}) to the observed Galactic GCLF requires a
conversion from luminosity to mass. 
For the FZ01 model ($\gamma = 1$), we simply adopt a 
mass-independent $\cM/L_V$ ratio. 
For the KPZ09 model ($\gamma = 0.7$), we use their relation between
$\cM/L_V$ and $L_V$ at an age of 12 Gyr.\footnote{This relation was shown
in Fig.\ 4 of KPZ09, and with more clarity, in \citet{krupor10}.}  
The corresponding parameters of the KPZ09 model are: \citet{king66}
concentration parameter $W_0 = 7$ (the median value of $W_0$ for the Milky Way
GC system, \citealt{harr96}), metallicity $Z = 0.0004$, and $t_0$ = 1.3
Myr. The last of these parameters is defined in the KPZ09 model as $t_0 
\equiv t_{\rm dis} \: (\cM/M_{\odot})^{-\gamma}$, i.e., the dissolution timescale for a
cluster with $\cM = 1\; M_{\odot}$. The conversion to our notation is $t_0 =
\gamma/\mu$  (see equation~\ref{eq:dMdt}).   

We derive the observed GCLF of the Milky Way from the 2010 version of the 
\citet{harr96} catalog of GC data. To avoid uncertain luminosities, we
exclude 17 GCs with ${\it E}_{{\it B}-{\it V}} > 1.5$, resulting
in a catalog with 140 GCs. $V$-band luminosities are derived by assuming the standard
Galactic reddening law (with $A_V = 3.1 \; {\it E}_{{\it B}-{\it V}}$) 
and a solar absolute $V$-band magnitude of $M_V^0 = 4.83$
\citep{binmer98}.   

For the single-$\mu$ case, we perform least-squares fits of
the model GCLFs to the observed Galactic GCLF using the
non-linear Marquardt-Levenberg algorithm \citep{leve44,marq63}. This is done
for $\gamma = 1$ to represent the FZ01 model and for $\gamma = 0.7$ to
represent the KPZ09 model. We adopt $\beta = -2$, as found in several studies
of young star cluster  systems (\citealt{falcha12} and references therein).  
Table~\ref{t:GCLFfits} lists the reduced $\chi^2$ values of the two model fits
along with the resulting values of the evaporation rate $\mu$ and the cutoff
and peak luminosities, $L_{\rm c}$ and $L_{\rm p}$, of the GCLF. Values for
$L_{\rm p}$ were derived using equation (7) of \citet{goud12}.   

\begin{table}[tbh]
\footnotesize
\caption[]{Fits of Single-$\mu$ Models to Galactic GCLF.}
\label{t:GCLFfits}
\smallskip
\begin{tabular}{@{}rcclcc@{}} \tableline \tableline
\multicolumn{3}{c}{~~} \\ [-2ex]  
 \multicolumn{1}{c}{Model} & log $L_{\rm p}/L_{V,\odot}\!\!\!\!$ & log $L_{\rm
   c}/L_{V,\odot}\!\!\!\!$ & \multicolumn{1}{c}{$(\mu\,t)^{1/\gamma}$} & rms$\!\!$ & $\chi^2_{\rm red}$ \\ 
\multicolumn{1}{c}{(1)} & (2) & (3) & \multicolumn{1}{c}{(4)} & (5) & (6) \\ [0.5ex] \tableline  
\multicolumn{3}{c}{~~} \\ [-1.5ex] 
    FZ01 & 5.24 $\pm$ 0.04$\!\!\!\!$ & 7.15 $\pm$ 0.05$\!\!\!\!\!$ & 
    $(1.80 \pm 0.03) \; 10^5 \times \Upsilon_V \!\!\!\!\!\!\!$ & 0.25$\!\!$ & 0.68 \\
     K09 & 5.25 $\pm$ 0.03$\!\!\!\!$ & 7.21 $\pm$ 0.04$\!\!\!\!\!$ & $(4.94 \pm 0.05) \; 10^5$ & 0.21$\!\!$ & 0.60 \\
Gaussian & 5.24 $\pm$ 0.04$\!\!\!\!$ & N/A & \multicolumn{1}{c}{N/A} & 0.80$\!\!$ & 5.14 \\ [0.5ex] \tableline
\multicolumn{3}{c}{~~} \\ [-1.2ex]
\end{tabular}                   
\tablecomments{Column (1): model being fit to GCLF. (2): log of turnover luminosity in solar
  units. (3): log of Schechter cutoff luminosity in solar units. (4):
  value of $(\mu\,t)^{1/\gamma}$ in $M_{\odot}$ ($\Upsilon_V \equiv \cM/L_V$). (5): rms of
  residuals of fit to GCLF. (6): reduced $\chi^2$ value of fit to
  GCLF. See the discussion in \S\,\ref{s:GCLFfits}.  
}
\end{table}

We also made fits to the observed GCLF derived from the 2003 version of the Harris
catalog as a check for consistency with KPZ09 and to estimate the magnitude of
systematic errors in the fitted parameter values. We find that the values for
both $\mu\,t$  and $L_{\rm p}$ agree to within 0.5\% between the two versions of
the Harris catalog, while the values for $L_{\rm c}$ come out $\sim$\,30\% lower
for the 2003 version. These differences do not affect any of our conclusions.  

For $\cM/L_V$ = 1.8, which is the overall mean value found from
dynamical data in Sect.\ \ref{s:dyn} below, the evaporation rates of
the best-fitting single-$\mu$ FZ01 and KPZ09 models match each
other within only 6\% at $\mu \sim 2.6 \times 10^4 \; M_{\odot}$ Gyr$^{-1}$. As 
discussed in detail in \citet{mclfal08}, the corresponding value of 
$t_{\rm dis}/t_{\rm rlx} \sim 10 \; (-\beta - 1)^{-1} \sim 10$ where $\beta$ is the
power-law slope in equation~(\ref{eq:schech}).\footnote{Note that $\beta$ in this
  paper is equal to $-\beta$ in \citet{mclfal08}.} This value of $t_{\rm
  dis}/t_{\rm rlx}$ is at the low end of the range typically
found in theoretical calculations of single-mass clusters ($t_{\rm dis}/t_{\rm
  rlx} \sim 10-40$). However, simulations of multi-mass clusters
have shown evaporation rates that are significantly larger than those of
single-mass clusters \citep[e.g.,][]{leegoo95}. Furthermore, evaporation rates
depend on the specific techniques, assumptions, and approximations used for
each simulation \citep[see,
e.g.,][]{gneost97,vesheg97,baumak03,prigne08,hegg14}.   

For the multiple-$\mu$ case, we proceed as follows. We compute FZ01 and KPZ09
models from equation~(\ref{eq:multi-mu-MF}), adopting the values of $M_c$ listed in
Table~\ref{t:GCLFfits} along with $\beta = -2$ as before.  
To compute the evaporation rates $\mu_i$ of individual clusters, we again
assume an age $t = 12$ Gyr, and we use the values of $\mu\,t$ in
Table~\ref{t:GCLFfits} for a cluster with the 
median density $\hat{\rho_{\rm h}}$. We then use the current half-mass density
of each Galactic GC to estimate its evaporation rate from $\mu_i = (\rho_{{\rm
    h},\,i}/\hat{\rho_{\rm h}})^{1/2}$ (see equation \ref{eq:mu}).  
Again, we adopt the 2010 version of the \citet{harr96} catalog, 
  $M/L_V = 1.8$ for $\gamma = 1.0$ (i.e., the FZ01 model), and
  the KPZ09 relation between $M/L_V$ and $L_V$ for $\gamma = 0.7$.    

\begin{figure*}[htb]
\centerline{\includegraphics[width=13cm]{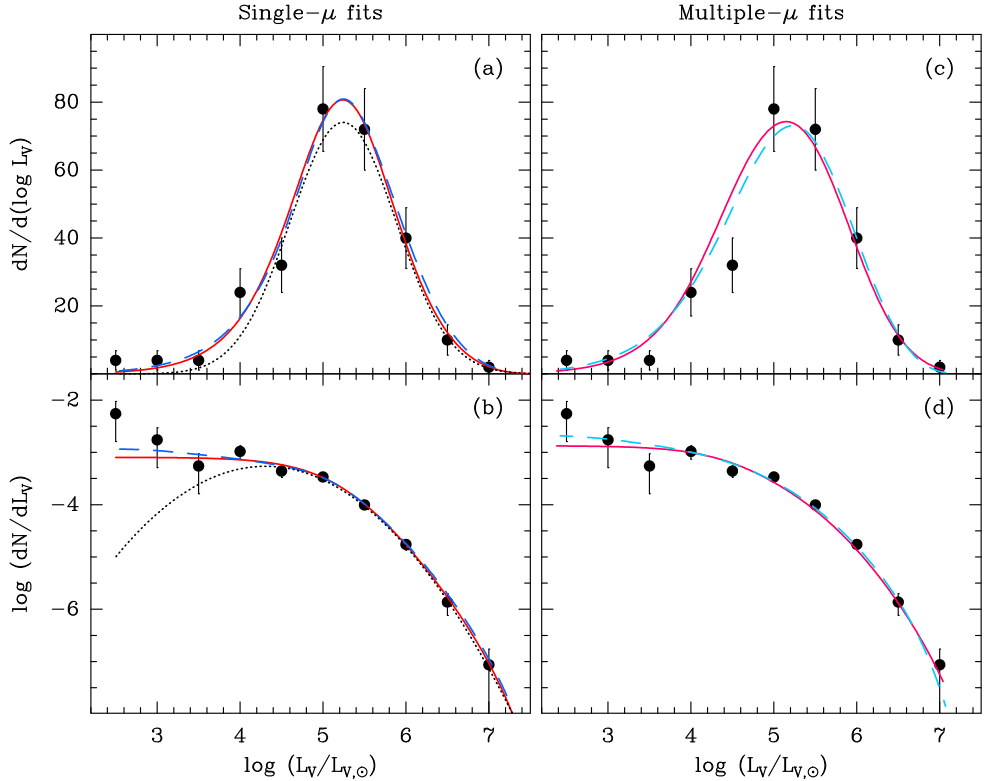}}
\caption{\emph{Panel (a)}: Fits of the single-$\mu$ FZ01 model (red solid
  curve), the single-$\mu$ KPZ09 model (blue dashed curve), and a Gaussian
  (black dotted curve) to the Galactic GCLF, expressed as the number of GCs per
  unit logarithm of $V$-band luminosity. 
  \emph{Panel (b)}: Similar to panel (a), but now in terms of $\log\,(dN/dL_V)$
  vs.\ $\log\,L_V$. 
  \emph{Panel (c)}: Similar to panel (a), but now for the multiple-$\mu$ models 
  described in the text. The light red solid curve represents the FZ01 model
  while the light blue dashed curve represents the KPZ09 model.  
  \emph{Panel (d)}: Similar to panel (c), but now in terms of $\log\,(dN/dL_V)$
  vs.\ $\log\,L_V$. All models are drawn for an age of 12 Gyr.  
}
\label{f:GCLFplot1}
\end{figure*}

The resulting FZ01 and KPZ09 models are compared with the observed
GCLF in Fig.~\ref{f:GCLFplot1}. Panels (a) and (c) show the GCLF in the form 
$dN/d\log L = (L\,\ln 10)\, dN/dL$ vs.\ $\log\,L$, analogous to 
the familiar observed GCLFs in magnitude space, whereas panels
  (b) and (d) show the GCLF in the form $\log\,(dN/dL)$ vs.\ $\log\,L$. 
For comparison with the more traditional log-normal representation of the
GCLF, we also overplot a best-fit Gaussian whose parameters are $\left<
  \log\,(L_V/L_{V\,\odot}) \right> = 5.24 \pm 0.04$ and $\sigma_{{\rm log}\,L}
= 0.64 \pm 0.05$ in panels (a) and (b). 

As Fig.~\ref{f:GCLFplot1} shows, the differences between the FZ01 and KPZ09 models
in their ability to fit the observed GCLF are very small relative to the
uncertainties. This holds for both the single-$\mu$ and 
  multiple-$\mu$ models. Statistically, the KPZ09 model fits the data slightly
better at $\log\,(L_V/L_{V,\,\odot}) \la 3.5$, while the FZ01 model fits better 
at $L \ga L_{\rm p}$. 
However, we emphasize that these differences are negligible not only with
respect to the poisson errors, but also when compared to the improvement that
the FZ01 and KPZ09 models provide over the traditional Gaussian representation
of the GCLF. This is due to the asymmetry in the observed GCLF in that
there are more GCs at $L\, < \, L_{\rm p}$ than at $L\, > \, L_{\rm p}$.
This asymmetry is especially clear in panels (b) and (d) of
Fig.~\ref{f:GCLFplot1} which show that 
$\psi\,(L) = dN/dL$ is approximately flat for GCs with $L_V \la
10^4\,L_{V,\,\odot}$. This behavior is clearly not consistent with a Gaussian
LF, while it is matched very well by both the FZ01 and the KPZ09 models.

Fig.~\ref{f:GCLFplot1} also shows that there is no significant difference
between the single-$\mu$ and multiple-$\mu$ models in terms of fitting the 
observed GCLFs, confirming the findings of FZ01 and KPZ09. Quantitatively, the
reduced $\chi^2$ values of the multiple-$\mu$ fits are 0.61 and 0.59 for the FZ01 and
KPZ09 models, respectively. In the remainder of this paper, we adopt the
single-$\mu$ models for simplicity.

It is worth noting that the GCLFs of the FZ01 and KPZ09 models are
more similar than are their GCMFs (see Fig.\ 1 of KPZ09). The
reason for this is that the different relations between $\cM/L_V$ and $L_V$
largely compensate for the differences in the GCMFs. 
We conclude that the shape of the Galactic GCLF does not provide 
any evidence for a GC mass-dependent $\cM/L_V$ ratio. Thus, the 
claim by KPZ09 that ``the match between the models and the observations [of the
Galactic GCLF] exists only for values of $\gamma \approx 0.7$'' is not correct.

\section{Searches for Dynamical \texorpdfstring{$\cM/L$}{M/L}
  Variations}  
\label{s:dyn}
 
The most reliable estimates of GC masses are those derived from stellar
kinematics, often referred to as ``dynamical masses''. 
KPZ09 argued that the dependence of $M/L$ on $L$ predicted by 
their model was supported by the dynamical masses compiled by \citet{mand+91}
who found a weak correlation between $M/L$ and $M$ due to a few high $M/L$ values
at $\log\, (M/M_{\odot}) \ga 5.5$. However, no such correlation appears in the
more homogeneous and larger compilation of $M/L$ values by \citet{mcla00}, which
superseded most of the \citeauthor{mand+91}\ results.  
With this in mind, we review recent measurements of dynamical $\cM/L$ ratios
of ancient GCs in the literature to search for any dependence on GC
luminosity similar to that predicted by the KPZ09 model. 
We test for a dependence of $M/L$ on GC luminosity rather than GC mass because
the uncertainties in GC luminosities are typically relatively small and
similar among clusters. In contrast, the uncertainties in GC masses
vary significantly among clusters, depending on measurement specifics  
(e.g., numbers of stars measured, radial coverage). Thus, the correlation
between measurement errors of $M/L$ and $\log\,L$ is much smaller than between
those of $M/L$ and $\log\,M$. 

We include measurements from five independent studies of Galactic GCs:  
\citet[][37 GCs]{mclvdm05}, \citet[][14 GCs]{luet+13}, \citet[][14
GCs]{zari+12,zari+13,zari+14}, \citet[][25 GCs]{kimm+15}, and
\citet[][15 GCs]{watk+15}. In addition, we use the large sample of 178
old GCs in the Andromeda galaxy by \citet{stra+11}.

\subsection{Limits on $\cM/L_V$ Variations in Milky Way GCs}
\label{sub:Galdata}

\begin{figure*}[htbp]
%\centerline{\includegraphics[width=\textwidth]{MLplot_6panels.eps}}
\centerline{\includegraphics[width=0.85\textwidth]{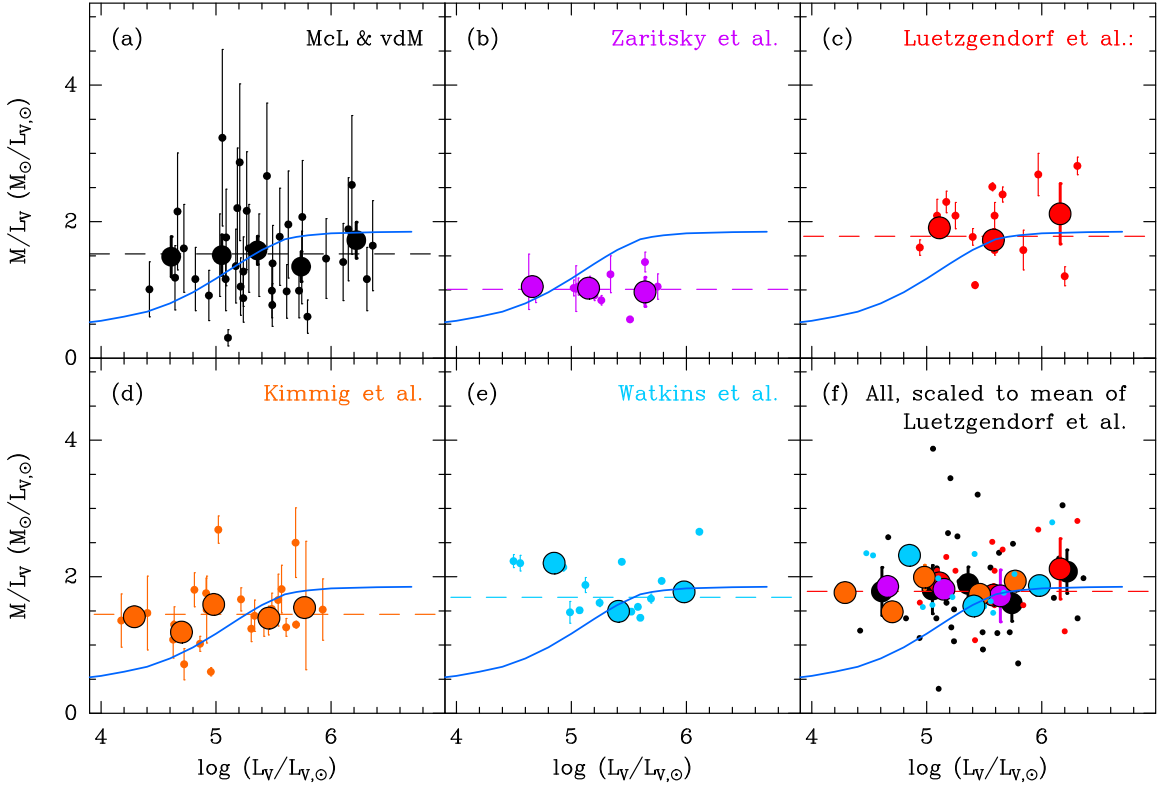}}
\caption{\emph{Panel (a)}: observed dynamical $\cM/L_V$ ratios of
  Galactic GCs from \citet[][black circles]{mclvdm05} as a function of their
  $V$-band luminosities. 
  The large filled circles depict average values in five bins of equal
  size in log $(L_V/L_{V,\,{\odot}})$. For comparison, we overplot the overall average
  value of $\cM/L_V$ (dashed line) and the relation between $\cM/L_V$ and log
  $(L_V/L_{V,\,{\odot}})$ at an age of 12 Gyr predicted by the KPZ09 model 
  (blue solid line). 
  \emph{Panel (b)}: similar to panel (a), but now showing data from
  \citet[][purple circles]{zari+12,zari+13,zari+14}. 
  \emph{Panel (c)}: similar to panel (a), but now showing data from 
  \citet[][red circles]{luet+13}. 
  \emph{Panel (d)}: similar to panel (a), but now showing data from 
  \citet[][orange circles]{kimm+15}. 
  \emph{Panel (e)}: similar to panel (a), but now showing data from 
  \citet[][light blue circles]{watk+15}. 
  \emph{Panel (f)}: a combination of all $\cM/L_V$ ratios from the five 
  studies depicted in panels (a)\,--\,(e), after correcting for systematic
  differences between the mean $\cM/L_V$ values of the individual studies (see
  text in Sect.\ \ref{sub:Galdata}.1). For clarity, the data for the 
  individual GCs in panel (f) are shown without errorbars.
}
\label{f:MLplot1}
\end{figure*}

The dynamical $\cM/L_V$ measurements of the five samples of Galactic GCs are
shown as a function of log $L_V$ in panels (a)\,--\,(e) of
Fig.~\ref{f:MLplot1}. 
For each cluster sample, the overall average value of $\cM/L_V$ is shown by a
horizontal dashed line. For comparison with the predictions of the KPZ09 
model, we also overplot their relation between $\cM/L_V$ and
$L_V$ at an age of 12 Gyr.

Systematic differences among dynamical $\cM/L_V$ values derived by the five
studies mentioned above are described in detail in the Appendix. Panel
(f) of Fig.\ \ref{f:MLplot1} shows our correction for these systematic
differences, which are mainly due to the different ways to convert observed
$\cM/L_V$ to ``global'' values that apply to each cluster as a whole. We
adopt the normalization of \citet{luet+13}. Thus, we multiply the
$\cM/L_V$ values of \citet{mclvdm05}, \citet{zari+12,zari+13,zari+14},
\citet{kimm+15}, and \citet{watk+15} by 
factors of 1.20, 1.75, 1.26, and 1.07, respectively (see Appendix~\ref{s:AppA}).   

\begin{table*}[tbh]
\footnotesize
\caption[]{Fits of Models to Observed Dynamical $\cM/L_V$ Values.}
\label{t:MLfits}
\begin{center}
\begin{tabular}{@{}lrrrrrrrrrrrr@{}} \tableline \tableline
\multicolumn{3}{c}{~~} \\ [-2ex]  
 & \multicolumn{2}{c}{McL \& vdM} &
 \multicolumn{2}{c}{L\"utzgendorf} & \multicolumn{2}{c}{Zaritsky} & 
 \multicolumn{2}{c}{Kimmig} & \multicolumn{2}{c}{Watkins} & 
 \multicolumn{2}{c}{Strader} \\
 Parameter & FZ01 & K09 &  FZ01 & K09 &  FZ01 & K09 &  FZ01 & K09 &  FZ01 & K09 &  
 FZ01 & K09 \\
\multicolumn{1}{c}{(1)} & (2) & (3) & (4) & (5) & (6) & (7) & (8) & (9) & 
 (10) & (11) & (12) & (13) \\ [0.5ex] \tableline  
\multicolumn{3}{c}{~~} \\ [-1.5ex] 
   $\sigma/\sqrt{N}$ & 0.73 & 0.81 & 0.57 & 0.61 & 0.27 & 0.49 & 0.56 & 0.72 &
                0.41 & 0.71 & 0.70 & 0.79 \\ 
 $\chi^2_{\rm red}$ & 0.38 & 0.49 & 0.29 & 0.32 & 0.04 & 0.10 & 0.21 & 0.33 &
                0.15 & 0.41 & 0.40 & 0.50 \\ [0.3ex] \tableline
\multicolumn{3}{c}{~~} \\ [-2.2ex] 
$\mid\!\left<\Delta\chi^2_{\rm red} \right>\!\mid$ &  \multicolumn{2}{c}{0.12} &
  \multicolumn{2}{c}{0.04} & \multicolumn{2}{c}{0.06} & \multicolumn{2}{c}{0.11} &
  \multicolumn{2}{c}{0.11} &  \multicolumn{2}{c}{0.08} \\ 
$\sigma(\mid\!\Delta \chi^2_{\rm red}\!\mid)$ & \multicolumn{2}{c}{0.08} &
  \multicolumn{2}{c}{0.05} & \multicolumn{2}{c}{0.05} & \multicolumn{2}{c}{0.06} &
  \multicolumn{2}{c}{0.03} &  \multicolumn{2}{c}{0.04}  \\ [0.5ex] \tableline
\multicolumn{3}{c}{~~} \\ [-1.8ex]
\end{tabular}                   
\tablecomments{Column (1): parameter name (standard error, reduced
  $\chi^2$, absolute mean value of $\Delta \chi^2_{\rm red}$, and standard
  deviation of $\mid\!\Delta \chi^2_{\rm red}\!\mid$). See the discussion in
  \S\,\ref{sub:Galdata} for the meaning of the latter two parameters.  
   Columns (2) and (3): values for FZ01 and K09 model fits to
  \citet{mclvdm05} data, respectively. 
  Columns (4) and (5): same as columns (2) and (3), respectively, but now for
  \citet{luet+13} data.  
  Columns (6) and (7): same as columns (2) and (3), respectively, but now for 
  \citet{zari+12,zari+13,zari+14} data.
  Columns (8) and (9): same as columns (2) and (3), respectively, but now for 
  \citet{kimm+15} data.
  Columns (10) and (11): same as columns (2) and (3), respectively, but now for 
  \citet{watk+15} data.
  Columns (12) and (13): same as columns (2) and (3), respectively, but now for 
  \citet{stra+11} data.
 }
\end{center}
\end{table*}

As Fig.~\ref{f:MLplot1} clearly shows, there is no evidence favoring the
relation between $\cM/L_V$ and $L_V$ in the KPZ09 model 
over the constant $\cM/L_V$ ratio in the FZ01 model.  
This is quantified in Table~\ref{t:MLfits}, which lists the $\chi^2$
values of the fits for the two models to the five data samples shown in 
Fig.~\ref{f:MLplot1}.\footnote{The fits of the KPZ09 model to the data were
  performed by allowing for a $L_V$-independent scale factor in
  $\cM/L_V$.} Specifically, the mass-independent $\cM/L_V$ model yields
somewhat better fits to the data of \citet{mclvdm05},
\citet{zari+12,zari+13,zari+14}, \citet{kimm+15}, and, more clearly, to 
the data of \citet{watk+15}. 

To put the $\chi^2$ values in Table~\ref{t:MLfits} in context, we performed a
series of Monte Carlo simulations as follows. For each of the five $M/L$ datasets with
individual luminosities $L_i$, we calculate masses $M_i$ under the assumptions
of both the FZ01 model (using a constant $M/L_V = 1.8$) and the KPZ09 model (using
their $M/L_V$ vs. $L_V$ relation). To each of the $L_i$ and $M_i$ values we then 
add random measurement errors based on the distribution of such errors in the
dataset in question. The synthetic $M_i/L_i$ data that are intrinsically
distributed like the FZ01 model are then fitted by the KPZ09 model, and vice
versa. We define $\Delta \chi_{\rm red}^2 \equiv \chi_{\rm
    red,\,KPZ09}^2 -  \chi_{\rm red,\,FZ01}^2$, i.e., the difference in
  $\chi_{\rm red}^2$ between the KPZ09 and FZ01 model fits. These simulations
were performed 1000 times for each of the five datasets. The resulting
distributions of $\Delta \chi_{\rm red}^2$ thus reflect the expected
probabilities to find a given difference in $\chi_{\rm red}^2$ between the
two model fits for the dataset in question. 

\begin{figure}[htbp]
%\centerline{\includegraphics[width=8.3cm]{chi2plot.eps}}
\centerline{\includegraphics[width=7.5cm]{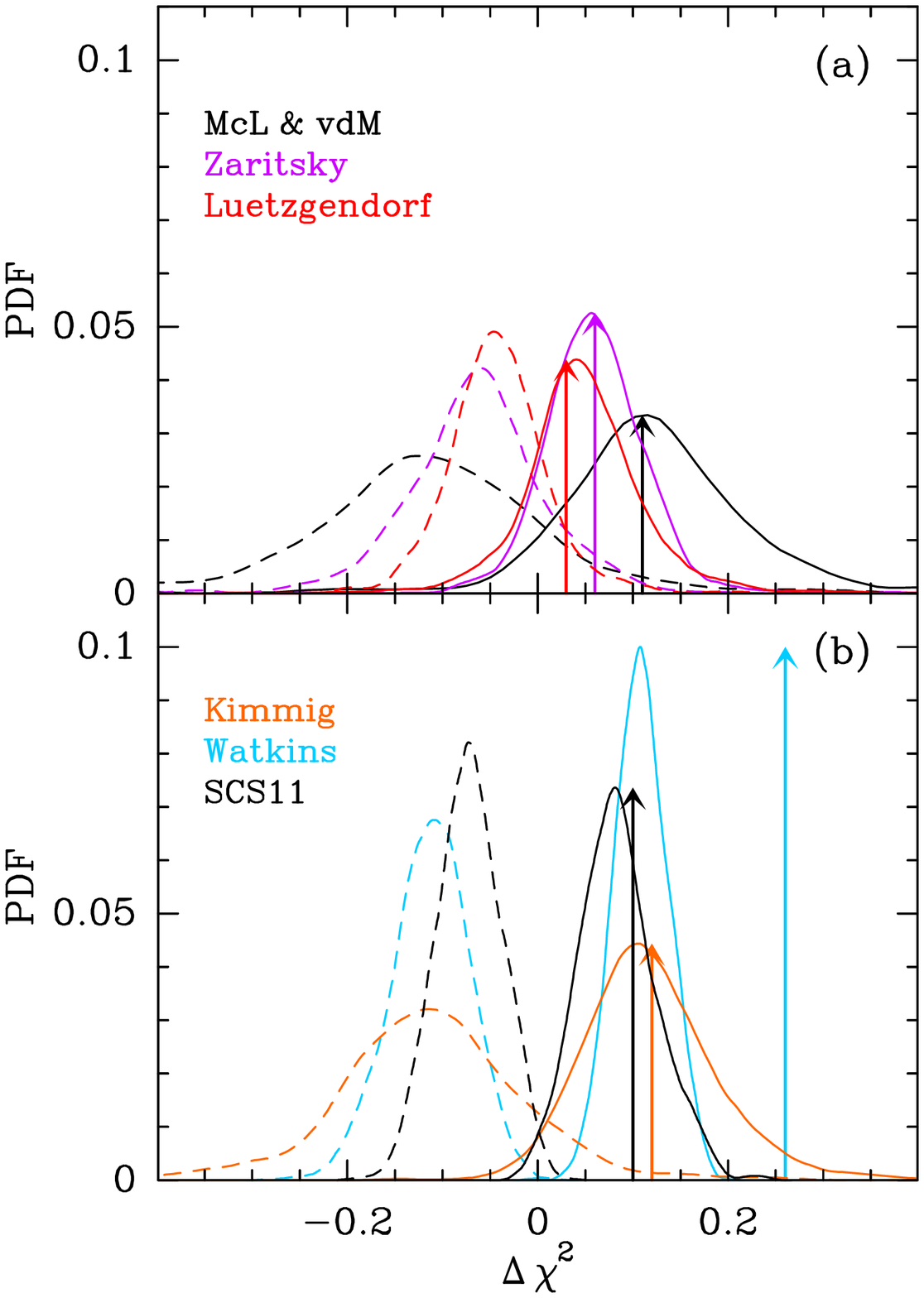}}
\caption{{\emph{Panel (a)}}: probability densities of 
  $\Delta \chi_{\rm red}^2$ values (see Sect.\ \ref{sub:Galdata} for its definition)
  for the $M/L_V$ datasets shown in Figs.\ \ref{f:MLplot1} and
  \ref{f:M31MLplot}: \citet[][black line]{mclvdm05}, 
  \citet[][purple line]{zari+12,zari+13,zari+14}, and 
  \citet[][red line]{luet+13}. For all datasets, the solid and dashed
    curves represent sets of Monte-Carlo simulations in which the $M/L_V$ values
    of GCs are intrinsically distributed according to the FZ01 and KPZ09 models,
    respectively.  
  For comparison, the measured differences in $\chi_{\rm red}^2$ between the
  KPZ09 and FZ01 model fits to the $M/L_V$ values of those datasets are shown
  by vertical solid arrows in the color of the dataset in question. 
  \emph{Panel (b)}: similar to panel (a), but now for the datasets of
  \citet[][orange line]{kimm+15}, \citet[][light blue
  line]{watk+15}, and SCS11 (black line; see Sect.\ \ref{sub:M31data})). 
}
\label{f:chi2plot}
\end{figure}

Fig.~\ref{f:chi2plot} shows the distributions of $\Delta \chi_{\rm red}^2$,
while Table~\ref{t:MLfits} lists the corresponding absolute mean values
$\mid\!\!\left< \Delta   \chi_{\rm red}^2 \right>\!\!\mid$ and standard
deviations $\sigma (\mid\!\!\Delta \chi_{\rm red}^2\!\!\mid)$. Comparing the
values of  $\left< \Delta \chi_{\rm red}^2 \right>$ with the measured
differences in $\chi_{\rm red}^2$ between the KPZ09 and FZ01 fits to
the five $M/L$ datasets (which are shown in Fig.~\ref{f:chi2plot} 
as vertical arrows), we find that the measured differences in $\chi_{\rm
  red}^2$ are all consistent with the hypothesis that a constant $M/L$ fits the 
data better than the KPZ09 model. Quantitatively, the KPZ09 model is excluded by
the data at confidence levels of 2.9\,$\sigma$, 1.4\,$\sigma$, 2.4\,$\sigma$,
3.8\,$\sigma$, and 12.3\,$\sigma$, for the datasets of
\citeauthor{mclvdm05}, \citeauthor{luet+13}, \citeauthor{zari+12},
\citeauthor{kimm+15}, and \citeauthor{watk+15}, respectively. 

\begin{figure}[htbp]
%\centerline{\includegraphics[width=8.cm]{ML_vs_FeH.eps}}
\centerline{\includegraphics[width=6.2cm]{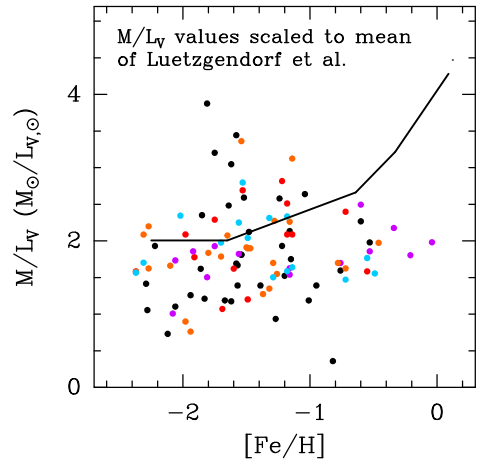}}
\caption{Measured $\cM/L_V$ values of Galactic GCs shown in panel (f) of Fig.\
  \ref{f:MLplot1} versus [Fe/H]. Symbol colors are the same as in Fig.\
  \ref{f:MLplot1}. For comparison, the solid line represents the SSP model
  predictions of \citet{bc03} for a \citet{chab03} IMF. 
}
\label{f:ML_vs_FeH}
\end{figure}

We also note the lack of a correlation between dynamical
$\cM/L_V$ and metallicity [Fe/H] at any GC luminosity in 
Figure~\ref{f:ML_vs_FeH}.  
This is not consistent with the relation predicted by simple stellar population
(SSP) models, as illustrated by the solid line in Fig.~\ref{f:ML_vs_FeH}. 
This suggests that the $M/L_V$ values of ancient GCs are more
  affected by their dynamical histories than by their
  metallicities.\footnote{This also explains why one can safely neglect the
  metallicity dependence of $\cM/L$ when fitting the GCLF by cluster evolution
  models as in Sect.~\ref{s:GCLFfits} (see also FZ01 and KPZ09).}  
For multi-mass King models of stellar systems, 
\citet{shagie15} showed that the lack of correlation between $\cM/L_V$ and
[Fe/H] among GCs can be explained by mass segregation, which causes the
brighter, more massive stars in the central regions, where the kinematic
measurements are typically made, to move with lower velocities than the
fainter, less massive stars in the outskirts, and whose effect is stronger at
higher metallicities due to the increasing turn-off mass with higher 
metallicity.  

The effect of stellar mass segregation on dynamical mass measurements is also
mass-depen\-dent, since present-day low-mass clusters have survived many more
relaxation times on average than high-mass clusters. This means that the
current level of mass segregation increases with decreasing GC mass, which
causes dynamical masses of GCs to be \emph{systematically underestimated for
  low-mass GCs}.\footnote{This trend might not extend to the lowest-mass
  clusters ($\log\,(M/M_{\odot}) \la 4.5$) which have only $\la 10\%$ of their 
  lifetimes remaining and may already have undergone core
  collapse (see, e.g., BM03).}    
  This reinforces our conclusion that the dynamical $\cM/L_V$ data show no
  evidence for the $\cM/L_V$ variations predicted by the KPZ09 model. 

\subsection{Limits on $\cM/L_V$ Variations in Andromeda GCs}
\label{sub:M31data}

We can also compare the model $M/L$ ratios to the observed ones
  for GCs in the Andromeda galaxy (M31). 
\citet[][hereafter SCS11]{stra+11} derived dynamical $\cM/L$
ratios for a large sample ($N = 178$) of old GCs in M31, covering a wide range
of luminosities ($4.7 \la \log\,(L_V/L_{V,\,\odot}) \la 6.5$). We adopt the GC
masses that they derived using the virial theorem, as well as
their $V$-band luminosities. We also follow SCS11 in discarding GCs whose
relative errors in $\cM/L_V$ are larger than 25\%.

To evaluate the luminosity dependence of the $\cM/L_V$ ratio from the M31 data,
we first consider a subselection in metallicity. As reported by SCS11, the
$\cM/L$ ratios of their metal-rich GCs ($\mbox{[Fe/H]} \ga -0.5$) are 
systematically and significantly lower than that of SSP model
predictions. This contrasts with the lower-metallicity GCs whose
$\cM/L_V$ ratios scatter around the SSP model predictions (cf.\ Fig.\ 1 of
SCS11). As mentioned in the previous subsection, this behavior
is consistent with mass segregation, whose effect is especially strong at
$\mbox{[Fe/H]} > -0.5$ \citep{shagie15}. To minimize the 
bias introduced by mass segregation (i.e., causing underestimates of $\cM/L$
whose amplitude is mass-dependent), we therefore subselect GCs with
$\mbox{[Fe/H]} < -0.5$.  

\begin{figure}[htbp]
%\centerline{\includegraphics[width=8.3cm]{M31_MLplot1.eps}}
\centerline{\includegraphics[width=6.2cm]{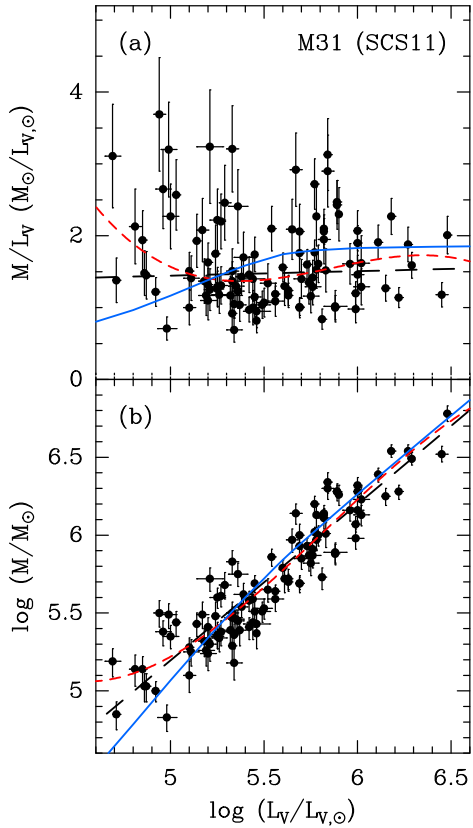}}
\caption{\emph{Panel (a)}: dynamical $\cM/L_V$ versus\ $\log\,L_V$ for 
  the sample of GCs in the Andromeda galaxy (M31) from SCS11. 
  For comparison, we overplot linear and cubic fits to the data 
  (long-dashed and short-dashed lines, respectively) and the relation between
  $\cM/L_V$ and log\,$L_V$ predicted by the KPZ09 model for an age of 12 Gyr
  (solid line). 
  \emph{Panel (b)}: similar to panel (a), but now for $\log\,\cM$ versus 
  $\log\,L_V$. See the discussion in Sect.\ \ref{sub:M31data}.
}
\label{f:M31MLplot}
\end{figure}

Panel (a) of Fig.\ \ref{f:M31MLplot} shows $\cM/L_V$ versus 
$\log\,L_V$ for the resulting sample of 109 GCs in M31. Linear and 
cubic fits to the data (using inverse variance weighting) are shown as black
and red dashed lines, respectively, while the prediction of the 
KPZ09 model is shown as a blue solid line. 
Overall, the picture is very similar to that for the Galactic GCs in
panel (c) of Fig.\ \ref{f:MLplot1} in that $\cM/L_V$ is again
  independent of $L_V$. Quantitatively, the linear fit to the data has a
slope $d(\cM/L_V)/d(\log\,L_V) = 0.06 \pm 0.13$, while the cubic fit actually 
shows a marginal upturn of $\cM/L_V$ at $\log\,(L_V/L_{V,\odot}) \la 5.2$,
where the KPZ09 model predicts a downturn. 

This result may seem surprising, since SCS11 reported a correlation between
$M/L_V$ and $\log\,M$ for the same dataset, similar to that predicted by the
KPZ09 model. We find that their apparent correlation results from a
strong covariance of $M/L_V$ and $\log\,M$, due to correlated errors, rather
than a true physical relationship.\footnote{A similar situation is found
  for the dataset of \citet{kimm+15} discussed in Sect.\ \ref{sub:Galdata}.}    
To illustrate this, we also perform linear and cubic fits
between the \emph{independent} variables $\log\,L_V$ and $\log\,\cM$ for the SCS11
dataset. A glance at panel (b) of Fig.\ \ref{f:M31MLplot} confirms the trends
seen in $\cM/L_V$ vs.\ $L_V$ discussed above.  
Finally, we calculate $\chi^2$ values of fits of the two models (i.e., constant
$\cM/L$ and the KPZ09 model) to the SCS11 data. These values are
listed in Table~\ref{t:MLfits}. Similar to our results from the Galactic GC
samples, we find that the constant $\cM/L$ model provides a better fit
to the M31 data than the KPZ09 model. 
We also performed Monte Carlo simulations like those described in Sect.\
  \ref{sub:Galdata} for the SCS11 dataset. The results are listed in
  Table~\ref{t:MLfits}. We find that the KPZ09 model is excluded by the SCS11
  data at a confidence level of 4.5\,$\sigma$. 

We conclude that the available data on dynamical $\cM/L_V$ ratios for ancient
GCs in the Milky Way and M31 provide no evidence for a dependence
on GC luminosity of the kind predicted by the KPZ09 model. Specifically, the
decline of $\cM/L_V$ with decreasing $L_V$ below $\log\,(L_V/L_{V,\,\odot})
\approx 5.2$ predicted by the KPZ09 model is not seen in the data.

\section{Constraints from Simultaneous fits to the GCLF and
  the $\cM/L_V$ versus $L_V$ Relation}
\label{s:Tdiss}

In this Section, we check whether the discrepancy between the observed
$\cM/L_V$ data at low GC luminosities and the predictions of the KPZ09
model discussed in Sect.~\ref{sub:Galdata} may be resolved by increasing the
characteristic dissolution timescale $t_0$, taking the fit to
the GCLF into account as well. To do this, we turn to the full set of K09
  models that are available online.\footnote{The K09 \texttt{SPACE} 
  models are available at \url{http://bit.ly/1Pbttlg}.} In the following, we
will refer to the KPZ09 model with $t_0 = 1.3$ Myr and $W_0 = 7$ discussed
above as the ``reference'' K09 model.  

\begin{figure}[htbp]
%\centerline{\includegraphics[width=7.8cm]{K09tests.eps}}
\centerline{\includegraphics[width=7.4cm]{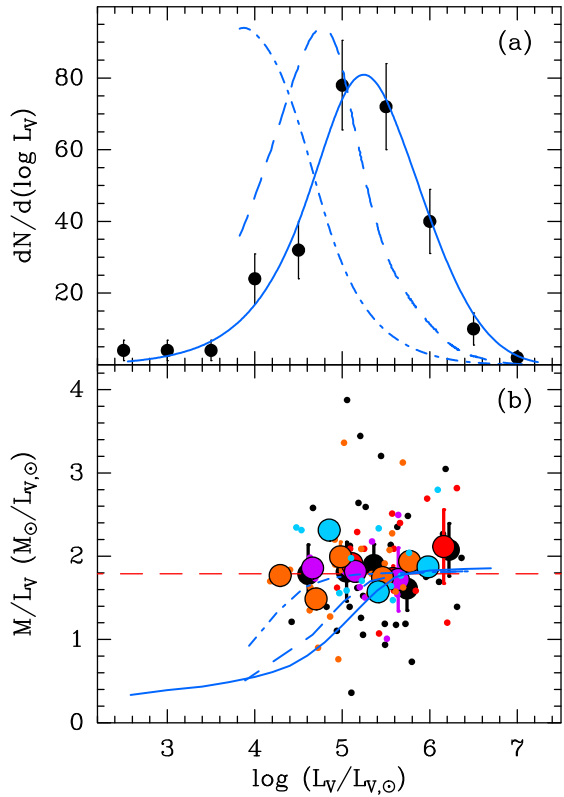}}
\caption{\emph{Panel (a)}: observed Galactic GCLF compared with the 
  K09 model at an age of 12 Gyr for three values of the dissolution timescale:
  $t_0$ = 1.3, 3, and 10 Myr (blue solid, dashed, and dashed-dotted lines,
  respectively). 
  \emph{Panel (b)}: dynamical $\cM/L_V$ ratios of Galactic GCs as a
  function of their $V$-band luminosities. This is a copy of panel (f) of 
  Fig.~\ref{f:MLplot1}, to which we have added K09 model predictions for $t_0$
  = 3 and 10 Myr in blue dashed and dash-dotted lines, respectively. 
  See the discussion in Section~\ref{s:Tdiss}.   
}
\label{f:K09tests}
\end{figure}

The impact of higher values of $t_0$ on the $\cM/L_V$ ratios predicted by the
K09 models as a function of $L_V$ is shown in panel (b) of
Fig.~\ref{f:K09tests}, which is a copy of panel (f) of Fig.~\ref{f:MLplot1} to
which we have added the K09 model predictions for $t_0$ = 3 and 10 Myr as dashed and
dash-dotted lines, respectively. Note that the K09 models with higher
values of $t_0$ are closer to the observed nearly flat distribution of dynamical
$\cM/L_V$ data than the reference model with $t_0$ = 1.3 Myr. 

We then calculate the corresponding GCLFs for an age of 12 Gyr predicted by the K09
models for $t_0$ = 3 and 10 Myr, respectively. 
To do so, we adopt the same initial GCMF as before, i.e., a Schechter
function with $\beta = -2$ and $M_{\rm c} = 9\times10^6 \; M_{\odot}$ (cf.\ Sect.\
\ref{s:GCLFfits} and Table~\ref{t:GCLFfits}). The initial GC masses are then
converted into masses and $V$-band luminosities at an age of 12 Gyr by means of
the K09 model tables, using spline interpolation.  
Finally, present-day GCLFs are calculated from the GCMFs using
$dN/d\log L_V = (dN/d\log M) \; (d\log M/d\log L_V)$  
where $d\log M/d\log L_V$ represents the local slope of the relation between
$\log L_V$ and $\log M$ at an age of 12 Gyr in the K09 models. 
The resulting GCLFs are depicted in panel (a) of Fig.~\ref{f:K09tests}.
Note that the GCLFs predicted for the K09 models with $t_0$ = 3 and 10 Myr
do not fit the GCLF well at all in that they peak at significantly lower 
luminosities than do the data and the reference K09 model (with $t_0 = 
1.3$ Myr), due to the lower evaporation rates. 
We have verified that the K09 model GCLFs for $t_0$ = 3 and 10 Myr are not
sensitive to the adopted cutoff mass $M_{\rm c}$ for $\log\,(L_V/L_{V,\,\odot})
\la 6$, even when $M_{\rm c}$ is increased by factors up to $10^3$.

We conclude that there is no value of $t_0$ for the K09 model that is able to
fit the GCLF and the $\cM/L_V$ data at log $(L_V/L_{V,\,\odot}) \la 5$ 
simultaneously. In contrast, the lack of a luminosity dependence of $\cM/L_V$
seen in panel (b) of Fig.~\ref{f:K09tests} is fitted naturally by the FZ01
model with a constant $\cM/L$. 

\begin{figure*}[tbh]
%\centerline{\includegraphics[width=12.cm]{relmuhist_epanech.eps}}
\centerline{\includegraphics[width=12.cm]{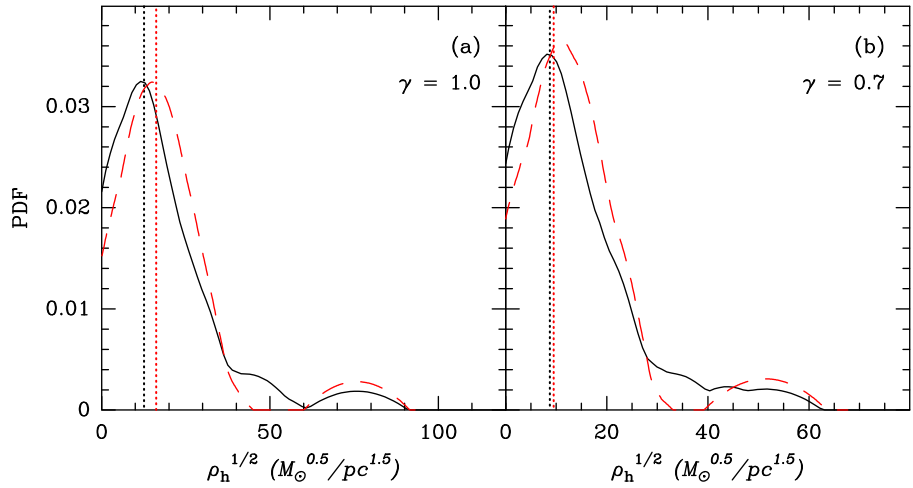}}
\caption{Probability densities of relative evaporation rates $\mu
  \propto \rho_{\rm h}^{1/2}$ of Galactic GCs. \emph{Panel (a)}: ``classical''
evaporation rates (for $\gamma = 1.0$). 
  \emph{Panel (b)}: ``retarded'' evaporation rates
  (for $\gamma = 0.7$). In both panels, the black solid line represents
  all Galactic GCs with $4.0 < \log\,(L_V/L_{V,\,\odot}) < 5.0$, while the red
  dashed line represents GCs with $4.0 < \log\,(L_V/L_{V,\,\odot}) < 5.0$ that
  have dynamical $\cM/L$ measurements shown in Figs.\ \ref{f:MLplot1} and
  \ref{f:K09tests}. Vertical black and red dotted lines indicate the median
  values of the respective distributions. See the discussion in
  Section~\ref{s:Tdiss}. 
}
\label{f:muhist}
\end{figure*}

To check the robustness of this conclusion, we compare the distributions of
relative evaporation rates $\mu$ of the GCs with log $(L_V/L_{V,\,\odot}) \leq 5$ 
that have dynamical $\cM/L_V$ measurements with those of all Milky Way GCs in
the same luminosity range %$4 < \log\,(L_V/L_{V,\,\odot}) < 5$ 
(cf.\  panel (b) of Fig.~\ref{f:K09tests}). Once again, we use the
2010 version of the \citet{harr96} catalog, using $M/L = 1.8$ for
$\gamma = 1.0$, and the KPZ09 $M/L_V$ vs.\ $L_V$ relation for $\gamma
= 0.7$ (cf.\ Section~\ref{sub:GCLFfit}).     
This comparison is shown in Fig.\ \ref{f:muhist}. Note
that the distributions of relative $\mu$ values of the two samples are very
similar to each other, with the median value actually being slightly larger for the
sample with dynamical $\cM/L_V$ measurements. This holds for both 
classical and retarded evaporation ($\gamma$ = 1.0 and 0.7). 
The inability of the K09 model to fit both the GCLF and the $\cM/L_V$ data
  at low luminosities simultaneously is therefore \emph{not} due to a mismatch
  between dissolution timescales of the low-luminosity Galactic GCs with
  available $\cM/L_V$ data and those of the full sample of Galactic GCs in the
  same luminosity range.

\section{Assessing Key Ingredients of the K09 Model}
\label{s:disc}

In this Section, we assess our findings from the preceding Sections in the
context of effects that were neglected in the FZ01 model but included in the
K09 model, and which KPZ09 claim are significant improvements.

\subsection{Mass Dependence of Retarded Evaporation Rates}
\label{sub:gamma}

We recall that the reference K09 model assumes $\gamma$ = 0.7 and $W_0 = 7$ for
clusters of all masses. This choice was based on approximations by
\citet{lame+10}, which in turn were based on the $N$-body simulations
by BM03 (cf.\ Sect.\ \ref{sub:GCMFs}). 
While retarded evaporation can be expected to occur in
all GCs at some level, it seems unlikely that a single value of 
$\gamma$ applies to GCs across the full range of initial masses. Because
  of the scaling $t_{\rm cr}/t_{\rm rlx} \propto M^{-1}$, stars that reach
escape velocities in high-mass GCs leave the cluster quicker relative to the 
situation in lower-mass GCs, implying that $\gamma$ should
increase with GC mass. In fact, if a fixed $\gamma < 1$ applied for all
GC masses, one would obtain an unphysical $t_{\rm dis} < t_{\rm rlx}$
at some high GC mass \citep[see also][]{baum01}. 
We therefore expect $\gamma$ to approach unity for GCs with sufficiently high
initial masses.  

\begin{figure}[tbh]
%\centerline{\includegraphics[width=7.5cm]{M12_vs_Minit.eps}}
\centerline{\includegraphics[width=6.5cm]{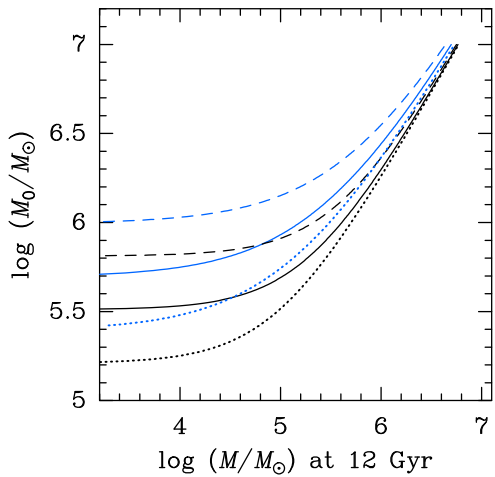}}
\caption{Relation between initial GC mass ($M_0$) and GC mass at an age of
  12 Gyr ($M$) for different dynamical evolution models. The solid lines show the
  single-$\mu$ models that fit the Galactic GCLF for $\gamma = 1.0$ (FZ01
  model; black line) and $\gamma = 0.7$ (KPZ09 model; blue line). The dashed
  and dotted lines in a given color represent the same models but with evaporation
  rates that are scaled up and down by a factor 2 relative to the best-fit
  values, respectively.}  
  See the discussion in Section~\ref{sub:gamma}.  
\label{f:massevol}
\end{figure}

The expected increase of $\gamma$ with initial cluster mass $M_0$ is relevant
because \citet{lame+10} derived the value $\gamma = 0.7$ from   
$N$-body simulations by BM03 with $4 \times 10^3 \la M_0/M_{\odot} \la 7
\times 10^4$ (corresponding to $8192 \leq N_0 \leq 131072$ for a \citet{krou01}
IMF). However,  
the great majority of GCs that survive for a Hubble time were initially
  much more massive than that. This is illustrated in Fig.~\ref{f:massevol}
which shows the relation between the masses at $t$ = 0 and 12 Gyr computed
  from equation~(\ref{eq:massevol}) for $\gamma$ = 1.0 and 0.7 with the best-fit
  values of $\mu$ from Table~\ref{t:GCLFfits}. For comparison, we also show the
  same models with evaporation rates that are factors 0.5 and 2.0 times those
  of the respective best-fit values.  
Note that the simulated clusters used to derive $\gamma = 0.7$ by
\citet{lame+10}, which have $M_0 \leq 10^{4.8}\; M_{\odot}$, do not even survive
12 Gyr of dynamical evolution according to these models. 
For a moderately low-mass GC with current mass $M \approx
10^{4.5}\;M_{\odot}$ for which the K09 models predict $\cM/L_V$ to be about
half of that of high-mass GCs (see Sect.~\ref{s:dyn}), the initial mass
indicated by these models is in the range $10^{5.5} - 10^6\;M_{\odot}$,
depending on the model. 
It is thus clear that \emph{the GCs that currently make up the bulk of the Galactic
GC system were initially at least one order of magnitude more massive 
than the simulated clusters used to derive 
$\gamma = 0.7$} by \citet{lame+10}. Since $\gamma$ is expected to increase with
increasing GC mass (cf.\ above), it seems prudent to regard the value $\gamma =
0.7$ adopted by K09 as a lower limit. 

In summary, $\gamma$ is expected to increase from $\approx 0.7$ for
  low-mass clusters to $\approx 1.0$ for high-mass clusters. 
Future $N$-body simulations with substantially more particles ($N$ in
the approximate range $10^{5.5}-10^{6.5}$ according to Fig.\ \ref{f:massevol})
will be needed to determine the actual dependence of $\gamma$ on 
  initial cluster mass.

\subsection{Other Assumptions in the K09 Model}
\label{sub:K09_other}

The semi-analytical model of K09 involves a large number of other parameters,
assumptions, and approximations (in addition to $\gamma = 0.7$ and $W_0
  = 7$ for the reference K09 model).   
Some of these ingredients are plausible, but some others are ad hoc
and/or not tested against observations, more rigorous theory, or
realistic $N$-body simulations. Given these uncertain inputs to the
model, it seems likely that the outputs from it will also be uncertain. 
Examples of assumptions in the K09 models whose quantitative effects are hard
to estimate include the following.  
\begin{itemize}
\item The initial-final mass relations for dark remnants (white dwarfs,
  neutron stars, and stellar-mass black holes).
\item The distributions of kick velocities of the various types of dark
  remnants, and the dependence of the retention fractions of such remnants on
  (initial) cluster escape velocity.  
\item The stellar mass dependence of the escape rate, for which the K09 models
  adopt the \citet{heno69} rate for \emph{close} stellar encounters in an
  \emph{isolated} cluster (i.e., not residing in a tidal field) without
  mass segregation. However, real GCs \emph{are} located in a (time-dependent)
  tidal field and stars escape mainly through repeated
  \emph{weak} encounters (i.e., two-body relaxation), for which the dependence
  of the escape rate on stellar mass may be different. 
\item The assumption of perfect energy equipartition during the
  mass-segregation phase of dynamical evolution. This assumption, which has an
  impact on the stellar mass dependence of evaporation, has recently been called
  into question since energy equipartition is not attained in $N$-body
  simulations except perhaps in their inner cores (BM03; 
  \citealt{trevdm13,soll+15,bian+16,sper+16}). 
\item The functional dependence of the stellar escape rate on the energy
  required for escape. 
\item The approximation of a cluster potential by a \citet{plum11} model, all
  the way out to the tidal radius (which does not exist for a Plummer
  model). 
\item An assumed relation between half-mass radius and initial cluster mass,
  specifically $r_{\rm h} \propto M^{0.1}$. Observed protoclusters,
  however, have a different relation, $r_{\rm h} \propto M^{0.4}$ \citep{fall+10}. 
\item The assumption that the half-mass radius $r_{\rm h}$ of a cluster remains
  constant throughout its lifetime. Recent $N$-body simulations show this to be
  an oversimplification. For example, $r_{\rm h}$ changes by a factor of
  $\sim 6$ during the lifetime of the $N$-body model for the globular
  cluster M4 by \citet{hegg14}. 
\end{itemize}
We refer the reader to the K09 paper for a full description and justification
of these and other ingredients of the K09 model.  

\begin{figure}[tbh]
%\centerline{\includegraphics[width=8.3cm]{ML_vs_time_BM03.eps}}
\centerline{\includegraphics[width=7.5cm]{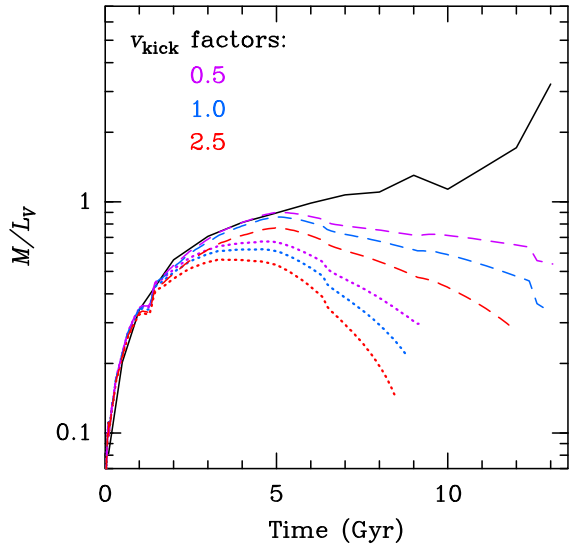}}
\caption{\emph{Black line}: evolution of $\cM/L_V$ in the $N$-body 
  simulation of BM03 for a $W_0 = 5$ cluster with initial mass $M_0 = 7 \times
  10^4 \; M_{\odot}$ %and apogalactic radius $R_A = 8.5$ kpc 
  and $t_0 = 10.7$ Myr discussed in
  Section~\ref{sub:K09_other}. \emph{Dashed lines}: K09 model for a $W_0 = 5$
  cluster with $M_0 = 10^5 \; M_{\odot}$, $t_0$ = 10 Myr, and three factors by
  which the ``standard'' kick velocities of dark remnants can be multiplied
  (0.5, 1.0, and 2.5 in purple, blue, and red, 
  respectively). \emph{Dotted lines}: same as dashed lines, but now for a
  cluster with $M_0 = 6 \times 10^4 \; M_{\odot}$. 
}
\label{f:MLcompare}
\end{figure}

As an illustration of uncertainties in the K09 model, we compare its predicted
$M/L$ evolution with that of a corresponding BM03 simulation.\footnote{The
properties of this BM03 simulation, shown in their Fig.\ 18, are: $W_0 = 5$ 
%, apogalactic radius $R_A = 8.5$ kpc, and orbit eccentricity
%$\epsilon = 0.5$ (see their Fig.\ 18). The corresponding 
and dissolution timescale $t_0 = 10.7$ Myr (see
equation 7 in \citealt{krumie09}). The properties of the corresponding K09
model are: $W_0 =  5$ and $t_0 = 10$ Myr.}  
We choose this particular comparison because the K09 model predictions
were normalized against the BM03 simulations. 
As shown in Fig.\ \ref{f:MLcompare}, 
the K09 model predictions follow the BM03 simulation quite well until an age of
a few Gyr, after which the $\cM/L_V$ decreases significantly in the K09 model
whereas it continues to \emph{increase} in the BM03 simulation, especially during the
last few Gyr of its lifetime of $\approx 14$ Gyr. The latter increase, which
implies an \emph{increasing $M/L$ with decreasing luminosity} at ages
$\ga 10$ Gyr, is thought to be due to the accumulation of massive white dwarfs
in the cluster (see BM03).  

The K09 model predictions were normalized against the BM03 simulations by using
the same initial conditions (see Sect.\ 4 in K09). However, the initial
conditions in the BM03 simulations differ from those in the published K09
models in one important aspect, namely the upper mass limit of the 
stellar IMF: BM03 used 15 $M_{\odot}$ (thus excluding progenitors of
stellar-mass black holes), whereas the K09 model uses 100 $M_{\odot}$, and 
includes prescriptions for retention fractions of, and kick velocities applied
by, neutron stars and stellar-mass black holes. Fig.\ \ref{f:MLcompare}
illustrates the significant effect of those kick velocities to $\cM/L_V$
in the K09 model.  

Note that the significant disagreement between $\cM/L_V$ in the BM03 simulation
and that in the K09 model predictions at ages $\ga 6$ Gyr exists for all
  choices of kick velocities. 
While the analytical implementation of kicks by dark remnants in the K09
model seems plausible at some level, there is significant uncertainty in the
retention fractions and kick velocities exerted by white dwarfs, neutron stars,
and black holes. Another related simplification in the K09 model is that it
applies a given retention fraction of dark remnants throughout the lifetime of
a cluster, while $N$-body simulations have shown that the retention fraction of 
stellar-mass black holes can decrease significantly during the lifetime of the
cluster due to multiple encounters
\citep{kulk+93,sigher93,pormcm00,merr+04,tren+10}. Furthermore, the  
evolution of the retention fraction of black holes varies widely among
repeated simulations with the same initial conditions \citep{merr+04}. It thus
seems fair to conclude that the $\cM/L_V$ decrease during the
second half of the lifetime of a cluster in the K09 models is not well
constrained by observations or simulations. 

We emphasize that this comparison between the K09 models and the BM03
simulation tests only a few of the many assumptions listed above. Most of the
others remain untested; however, it seems likely that they would also have some
impact on the resulting $\cM/L$ vs.\ time and $\cM/L$ vs.\ $M$ and $L$ relations. 

\subsection{Ambiguous Evidence from Stellar Mass Functions}
\label{sub:alpha}

One argument in favor of variations in cluster $M/L$ ratios 
is that observed stellar MF slopes $\alpha$ tend
to be relatively flat for low-luminosity GCs when compared with those for
high-luminosity GCs (\citealt{krumie09}; KPZ09). 
We re-examine this argument in this Section by comparing available
high-quality data on $\alpha$ in Galactic GCs with the K09 model predictions.

\begin{figure}[tbh]
%\centerline{\includegraphics[width=8.3cm]{alphaplot1_K09.eps}}
\centerline{\includegraphics[width=6.5cm]{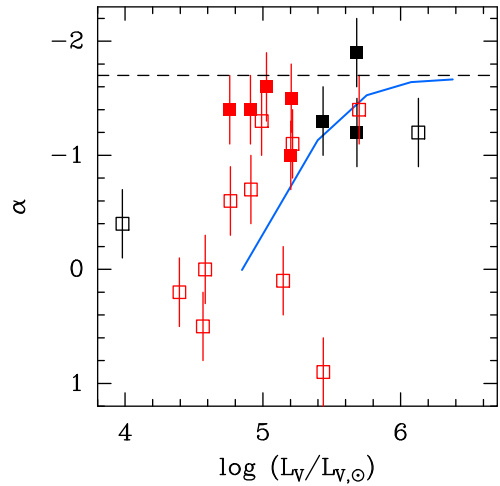}}
\caption{Stellar mass function slope $\alpha$ versus log\,$L_V$ for GCs in the
  sample of \citet{dema+07}. Filled squares and open squares
  represent GCs with King concentration indices $c > 1.4$ and $c < 1.4$
  (corresponding to $W_0 > 6$ and $W_0 < 6$), respectively. GCs with current
  half-mass relaxation times $<$\,1 Gyr are shown in red while the others are
  shown in black.   
  The solid blue curve represents the predictions of the reference K09 model. 
  The dashed line indicates $\alpha = -1.7$, the mean slope of the Kroupa IMF
  in the stellar mass range 0.3\,--\,0.8 $M_{\odot}$. See the 
  discussion in Sect.~\ref{sub:alpha}.  
  }
\label{f:alphaplot1}
\end{figure}

Fig.~\ref{f:alphaplot1} shows $\alpha$ for a mass
function $dn/dm \propto m^{\alpha}$ versus log $L_V$ from 
\citet{dema+07}, who compiled global MF slopes of 20 GCs in the stellar mass
range of 0.3\,--\,0.8 $M_{\odot}$ derived from \emph{Hubble Space
  Telescope (HST)} data. 
We assume measurement uncertainties of 0.3 dex for $\alpha$ (G.\ de Marchi,
private communication). The observed values of $\alpha$ are compared with
predictions of the reference K09 model in Fig.\ \ref{f:alphaplot1}. 

As mentioned by KPZ09, GCs in the fainter half of the sample studied by
\citet{dema+07} are on average more depleted in low-mass stars than those in the
brighter half (see Fig.\ \ref{f:alphaplot1}). This trend is, in principle, roughly
consistent with the predictions of the reference K09 model. 
However, it should also be noted that \emph{all} GCs that are
significantly depleted in low-mass stars relative to the most luminous
GCs feature relatively low King concentration parameters ($c \equiv
\log\,(r_{\rm t}/r_{\rm c}) \la 1.4$, corresponding to $W_0 \la
6$).\footnote{This tendency was already noted by \citet{dema+07} in
  a diagram of $\alpha$ versus $c$ (their Fig.\ 1). Our
  Fig.~\ref{f:alphaplot1} illustrates the luminosity dependence of the
  difference in $\alpha$ between GCs with high and low values of $c$.}   
Conversely, all GCs with $W_0 \ga 6$  
show $\alpha$ values consistent with a \citet{krou01} IMF, and thus do
not show any significant trend of $\cM/L_V$ with $L_V$. The latter does
not seem consistent with the predictions of the reference K09 model for the
Galactic GC system as a whole, which adopted $W_0 = 7$.  

However, the strong depletions of low-mass stars seen in GCs with low
concentration indices can also be explained in a different way:   
\citet{baum+08b} argued that this observation may be caused by 
low-concentration GCs having started out as tidally limited clusters with
relatively high levels of primordial mass segregation. The low-mass stars in
such clusters would initially be located relatively close to the
tidal radius, allowing their evaporation from the cluster to start well before
the cluster experiences core collapse, thus leading to present-day mass
functions that are relatively strongly depleted in low-mass stars. 
In this context, we also note that the $N$-body simulations by \citet{tren+10}
showed that GCs with low initial concentration index ($W_0$ = 5 or 3) gradually
evolve to higher concentration indices ($W_0 \sim 7$) within 40\,--\,70\% of
their dissolution time. As such, low-luminosity GCs with \emph{current} low
concentration indices likely had even lower concentration indices initially,
which might have caused the strong depletion of low-mass stars in such GCs to
start even earlier than predicted by the \citet{baum+08b} study. 
This effect is not incorporated in the K09 models, rendering it hard to
determine the reason for the significant difference in $\alpha$ between GCs with
low and high concentration indices without additional information. 

In conclusion, the observed depletion of low-mass stars in low-luminosity GCs 
can be explained in more than one way and therefore is a weak or inconclusive
test of the stellar mass dependence of the escape rate in the K09 models. 
Hence our conclusions from the previous sections remain valid.

\section{Summary and Conclusions}
\label{s:conc}
The most promising explanation for the peaked shape of the observed luminosity
function (LF) of old GCs is that it is a relic of dynamical
processes---primarily stellar escape driven by internal two-body
relaxation---operating on an initial power-law or Schechter MF of
young clusters over a Hubble time. The semi-analytical models of FZ01 
showed quantitative agreement with the present-day observed GCLF for a wide range of
initial MF shapes. For the sake of simplicity, and based on theoretical and
observational standard practice at the time, FZ01 adopted an evaporation rate
independent of cluster mass and a mass-to-light ratio
independent of cluster mass.  
KPZ09 challenged both of these assumptions. They claimed that the
predicted GCLF was only consistent with the observed GCLF if the stellar
evaporation rate depends significantly on cluster mass: $dM/dt \propto
M^{1-\gamma}$ with $\gamma = 0.7$ rather than $\gamma = 1$ for the FZ01
model. To calculate the escape rates of stars of different masses and hence the
variation of $\cM/L$ among clusters with different masses, they employed the
semi-analytical model of K09, which involves a significant 
number of plausible, but largely untested, assumptions and approximations.  

In this paper, we performed a quantitative evaluation of the KPZ09 claim that
their model could fit the observed GCLF while the FZ01 model could not. We
conclude that this claim is not valid, based on the following analysis and
results.  
\begin{enumerate}
\item 
  The FZ01 and KPZ09 models provide equally good fits to the observed GCLF in
  the Milky Way. Furthermore, both models yield a significantly
  better fit to the observed GCLF at low luminosities than the traditional
  Gaussian model, highlighting the importance of mass loss driven by
    two-body relaxation in shaping the GCLF. 
\item 
  The measured $\cM/L_V$ values of GCs in the Milky Way and the Andromeda
  galaxy show no dependence on cluster luminosity. At low GC luminosities
  ($\log\,(L_V/L_{V,\,\odot}) \la 5$), where the impact of a stellar
  mass-dependent escape rate is expected to be strongest, the observations are
  fitted better by a mass-independent $\cM/L_V$ than by the KPZ09 model. This
  result holds for all six independent studies of GCs with dynamical $\cM/L_V$
  data analyzed here.  
\item 
  We find that the discrepancy between the observed $\cM/L_V$ data at low GC
  luminosities and the KPZ09 predictions cannot be resolved by increasing
  the characteristic dissolution timescale $t_0$ of the K09 model, since such
  an increase would yield an unacceptable fit to the GCLF. In other words,
  there is no value of $t_0$ that allows the K09 model to fit simultaneously
  the GCLF and the observed $\cM/L_V$ data at $\log\,(L_V/L_{V,\,\odot}) \la 5$. 
\item
  The parameter $\gamma = 0.7$ adopted by KPZ09 is 
  based on results of $N$-body simulations of GCs with initial masses $M_0 \la
  7\times 10^4\; M_{\odot}$, whereas the initial masses of GCs that survive 12 Gyr
  of dynamical evolution are at least one order of magnitude
  higher than that. Theory indicates that the value of $\gamma$ will 
  increase toward unity at higher masses. Thus, the appropriate
  value of $\gamma$ for models of the GCMF and GCLF evolution may be
  closer to 1.0 than to 0.7. 
\end{enumerate}

We emphasize again that we do not dispute the physical principles of retarded
evaporation and $M/L$ variations. Rather, we claim that these effects add 
substantially to the complexity of dynamical GCMF and GCLF models and are not
needed in practice to match observed GCLFs.  

\acknowledgments
We thank Laura Watkins and the anonymous referee for helpful comments. This
project was partially supported by \emph{HST} Program GO-11691 which was
provided by NASA through a grant from the Space Telescope Science Institute,
which is operated by the Association of Universities for Research in Astronomy,
Inc., under NASA contract NAS5--26555. We acknowledge the use of the $R$
Language for Statistical Computing, see \url{http://www.R-project.org}.

%Facilities: \facility{HST (ACS)}, \facility{Keck:I (LRIS)}, \facility{ESO:3.6m
%  (EFOSC2)}

\appendix

\section{Systematic Differences between Dynamical $M/L$ Studies}
\label{s:AppA}

In this Appendix, we analyze and quantify systematic differences between
the five sources of dynamical $\cM/L_V$ measurements used in
Section~\ref{sub:Galdata} so that they can be combined in a
useful way.  
The $\cM/L_V$ values from \citet{mclvdm05} were derived from central velocity
dispersions from \citet{prymey93} which were then extrapolated to ``global''
values (for the cluster as a whole) using surface brightness
profiles. \citet{mclvdm05} used single-mass King models in this 
extrapolation, so that any radial gradients of $\cM/L_V$ are neglected. Since
ancient GCs commonly display radial mass segregation
\citep[e.g.,][]{meyheg97}, which causes the more massive stars to be more
centrally concentrated than the less massive stars (which have 
  higher $\cM/L$), we treat $\cM/L_V$ values from \citet{mclvdm05} as lower
limits.  

\citet{zari+12,zari+13,zari+14} measured velocity dispersions using a
drift-scan technique that moved the spectrograph slit across the target
cluster during the exposures, covering roughly the region within the half-light
radius. The $\cM/L_V$ ratios of Zaritsky et al.\ were determined using an
empirical relation between the half-light radius, the average surface
brightness within that radius, and the mass-to-light ratio within that
radius. This scaling relation was found to apply to all stellar systems from
star clusters to massive elliptical galaxies. However, as discussed in
\citet{zari+12}, their method produces $\cM/L_V$ values that are on average
$\sim 40-50$\% lower than those of \citet{mclvdm05}. This is consistent with the
observation that ancient star clusters lie systematically somewhat above the
empirical relation used by Zaritsky et al.\ \citep[see Fig.\ 2 in][]{zari+11}. 

The kinematic data analyzed by \citet{kimm+15} consisted of radial velocities of
individual cluster stars, both from new observations and from the
literature. GC masses were determined by fitting single-mass King models to
the observed radial velocity dispersion profiles. Similar to the case of
\citet{mclvdm05}, we thus treat the $\cM/L_V$ values from \citet{kimm+15} as
lower limits.     

\citet{watk+15} derived $\cM/L_V$ ratios by fitting dynamical models to a
combination of proper-motion velocity dispersions (from multi-epoch
\emph{HST} imaging data) and spectroscopic line-of-sight velocity
dispersions. Their fitting involved Jeans models that assume a constant $\cM/L$
ratio, which we therefore formally treat as lower limits. However, the
dispersion data used by \citet{watk+15} covered a large range of radii, and no
assumptions were made regarding the radial luminosity density profile, since they
used a Multi-Gaussian Expansion fit to the latter. Hence, their resulting
$\cM/L$ values can be expected to represent the cluster as a whole relatively
well. 

Finally, the integrated-light kinematics in \citet{luet+13} were derived from 
integral-field spectroscopy with spatial coverage typically out to the 
half-light radii of the clusters. Along with surface brightness profiles
derived from \emph{HST} data, the $\cM/L_V$ values in \citet{luet+13} were
determined using  Jeans modeling. Their method incorporates a correction for
radially varying $\cM/L_V$ and as such seems likely to produce results that
are more robust relative to mass segregation than the other studies
mentioned above.  
From the 11 GCs in common between the studies of \citet{luet+13} and 
\citet{mclvdm05}, the ratio of the $\cM/L_V$ values is 1.20 
$\pm$ 0.10 where the quoted uncertainty is the standard error of the
mean. Similarly, the mean ratio of the $\cM/L_V$ values of \citet{luet+13} and 
those of \citet{kimm+15} is 1.26 $\pm$ 0.25 for the 4 GCs in common between the
two studies, whereas that ratio is 1.07 $\pm$ 0.10 for the 7 GCs in common
between \citet{luet+13} and \citet{watk+15}. 
For the purposes of this paper, we suggest that these ratios are useful
estimates of the factor by which $\cM/L_V$ values may be systematically
underestimated in the studies that assumed a constant $\cM/L$ throughout the
cluster.\footnote{As mentioned in Sect.\ \ref{sub:Galdata}, the level
  of mass segregation is expected to depend on cluster mass to some
  extent. We neglect this effect, which is likely most significant for studies that
  use central velocity dispersions (e.g., \citealt{mclvdm05}).} 

Panel (f) of Fig.\ \ref{f:MLplot1} depicts our corrections for the
systematic differences between the $\cM/L_V$ estimates of the five 
studies described above. We adopt the normalization of \citet{luet+13}. 
For consistency with this normalization, we multiplied the $\cM/L_V$
values of \citet{mclvdm05}, \citet{zari+12,zari+13,zari+14}, \citet{kimm+15}, 
and \citet{watk+15} by factors of 1.20, 1.75, 1.26, and 1.07,
respectively. 

\end{document}